\documentclass[twocolumn]{aa}
\usepackage{graphicx}
\usepackage{txfonts}
\begin{document}
   \title{Atmospheric Turbulence Compensation with Laser Phase Shifting Interferometry}

   \author{Sebastian Rabien\inst{1}
   \and
   Frank Eisenhauer\inst{1}
   \and
   Reinhard Genzel\inst{1,2}
   \and
   Richard I. Davies\inst{1}
   \and
   Thomas Ott\inst{1}
   }

   \institute{Max-Planck-Institut f\"ur extraterrestrische Physik, Giessenbachstrasse, D-85748 Garching\\
              \email{srabien@mpe.mpg.de}\\
              \and Department of Physics, University of California, Berkeley, USA}

\authorrunning{Rabien et al.}
\titlerunning{Atmospheric Turbulence Compensation with Laser Phase Shifting
  Interferometry}

\abstract{Laser guide stars with adaptive optics allow astronomical
   image correction in the absence of a natural guide star. Single
   guide star systems with a star created in the earth's sodium layer
   can be used to
   correct the wavefront in the near infrared spectral regime for 8-m
   class telescopes. For
possible future telescopes of larger sizes, or for correction at
   shorter wavelengths, the use of a single guide star is ultimately
   limited by focal
anisoplanatism that arises from the finite height of the guide
   star. To overcome this limitation we propose to overlap coherently
   pulsed laser beams that are
expanded over the full aperture of the telescope, traveling upwards
   along the same path which light from the astronomical object
   travels downwards. Imaging the
scattered light from the resultant interference pattern with a camera
   gated to a certain height above the telescope, and using phase
   shifting interferometry we
have found a method to retrieve the local wavefront gradients. By
   sensing the backscattered light from two different heights, one can
   fully remove the cone
effect, which can otherwise be a serious handicap to the use of laser
   guide stars at shorter wavelengths or on larger telescopes. Using
   two laser beams multiconjugate correction is possible, resulting in
   larger corrected fields. With a proper choice of laser,
   wavefront correction could be expanded to the
visible regime and, due to the lack of a cone effect, the method is
   applicable to any size of telescope. Finally the position of the
   laser spot could be imaged
from the side of the main telescope against a bright background star
   to retrieve tip-tilt information, which would greatly improve the
   sky coverage of the
system. \keywords{telescopes -- instrumentation: adaptive optics --
   techniques: high angular resolution} }

   \maketitle

\section{Introduction}

Adaptive optics correction of wavefront distortions is becoming a common tool at large ground based telescopes. Diffraction limited imaging through the
turbulent atmosphere can be regarded as routine at near infrared wavelengths and shows great success in many scientific areas (see for example the proceedings
of the `Science with adaptive optics' meeting, eds. Brandner \& Kasper \cite{Brandner05}). By using laser guide stars this technique can be applied at almost
any position in the sky where a suitable star for tip-tilt correction can be found. Nevertheless, the quality of correction when using a single guide star is
limited by focal anisoplanatism, or the `cone effect'. Due to the limited height where the artificial guide star is created, some part of the wavefront
distortion remains unsampled, and thus uncorrected. For possible future extremely large telescopes, or for correction at short wavelengths, this limit will
ultimately prevent diffraction limited performance. To overcome this, tomographic solutions have been proposed (e.g. by Angel \& Lloyd-Hart \cite{Angel00a}),
utilizing multiple Rayleigh guide stars or multiple stars in the earth's sodium layer. For a large telescope or for correction in the visible a large number of
guide stars and wavefront sensors are needed.

Another approach is to use the upward path of a projected laser beam to sample the atmospheric turbulence, either in an inverse Shack-Hartmann geometry by
distributing many laser beams in parallel over the telescope, or with a single laser beam expanded over the whole aperture of the telescope. Buscher et al.
\cite{Buscher2002} have proposed a method to sample the local curvature of the atmospheric wavefront distortions from the intensity changes that a collimated
laser beam shows when it has traveled to a certain height. In the same publication a method is proposed to compare the full aperture expanded beam with lots of
other laser beams in a point-diffraction scheme. Also using a full aperture expanded laser, Bonaccini et al. \cite{Bonaccini2004} have proposed to measure the
wavefront on the downward path when imaging back the distributed beacon with a shearing interferometer. Feasibility studies of all the above mentioned methods
are currently being carried out. Limitations for the inverse Shack-Hartmann and the point-diffraction scheme may arise due to both the size of the individual
spots that can be reached in the presence of diffraction and also beam quality aspects. For the shearing measurement, the coherence length of the scattered
light might be a serious issue.

The method proposed here uses the coherent superposition of tilted laser wavefronts over the whole aperture of the telescope together with the method of laser
phase shifting interferometry (LPSI) in order to retrieve a local phase difference that can be used in an adaptive optics system in a similar way to the
gradients retrieved from a standard Hartmann sensor. To adapt to different conditions, the sensing resolution can be chosen freely by adding an appropriate
tilt to the beams. Detecting the light scattered from low altitudes will result in a ground layer correction. By extending the system to multiple detections at
different heights (or two beam systems) with more deformable mirrors, a complete sampling of the turbulent layers in the atmosphere is possible. Therefore high
Strehl ratios can be reached and multiconjugate adaptive optics correction over an extended field of view is possible. Additional means can help to expand the
sky coverage further, since in a differential measurement far off-axis guide stars can serve as the tip-tilt reference.

\begin{figure}
\centering \includegraphics[width=8cm]{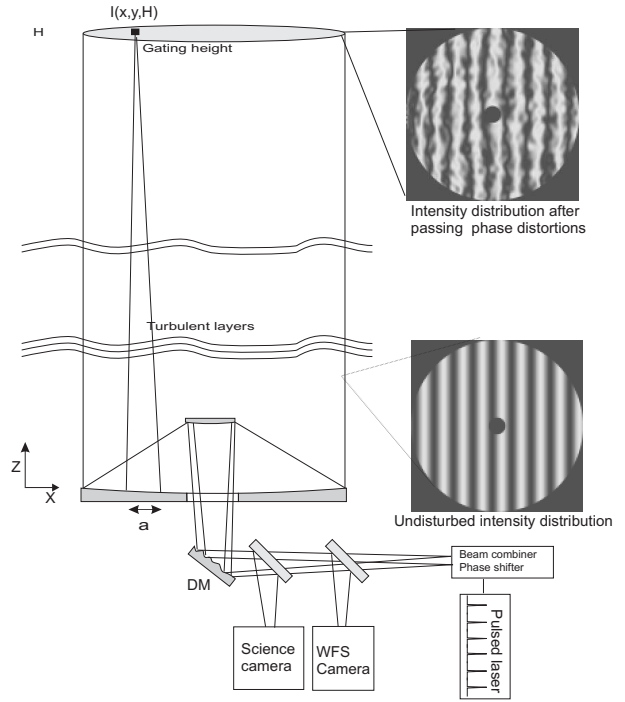} \caption{Basic principle of measuring atmospheric turbulence with
  phase shifting interferometry.
A pulsed laser beam is split into two beams which are tilted with respect to
  each other. To one of the beams a known phase shift $\alpha$ can be
  applied. The scattered
fringe pattern which is formed in the atmosphere at height H, can be
  viewed with a camera over the full telescope aperture, which is
  gated according to the time of flight. In a series of laser shots a
  known phase difference is added to the beam in each pulse. Depending
  on the algorithm used for the phase retrieval, a
set of three or more laser shots will allow the local phase difference
  to be measured. The frames shown here of a single laser shot are taken
  from a simulation with multiple turbulence layers of the fringe
  formation and imaging at an 8-m telescope with
  $r_0=10$\,cm.}\label{fig:basic_principle}
\end{figure}

\section{On-sky phase shifting interferometry}

Widely used for the measurement of optical surfaces, phase shifting interferometry (PSI) allows a phase difference to be measured with high accuracy. Standard
laboratory interferometers make use of the comparison between a reference phase and a phase under test, e.g. in a Twyman-Green setup. The reference beam is
phase shifted in several steps, and an interferogram is recorded at each step. Depending on the number of phase steps, several algorithms are possible to
calculate the phase difference and these can be found in textbooks (e.g. Malacara \cite{Malacara92}). For atmospheric turbulence probing the usual reference
flat is of course not present, which means that a global wavefront can not be measured directly.

Instead, with the scheme shown in Fig.~\ref{fig:basic_principle}, a local phase difference can be measured that allows an interpretation as local gradients,
and which can be treated as such for the wavefront correction. Assuming one pulse with two flat coherent laser wavefronts leaving the telescope --- one
slightly tilted with respect to the other --- an interference pattern can be created at any distance. When reaching a certain height in the atmosphere, both
waves will have accumulated local phase changes. Due to the tilt in the wavefront, the path through the atmosphere to the point (x,y,H) of each wave is
slightly different. The lateral shear of the waves at ground level is denoted by $a$ in Fig.~\ref{fig:basic_principle}. This creates a phase difference
$\Phi_a$ which modulates the intensity of the interferogram that is written in the sky. With a camera gated to twice the time of flight to the point (x,y,H)
the scattered light from this pattern can be imaged using the full telescope aperture. It should be noted that the large wavefront distortions present in the
atmosphere result only in moderate gradients to be probed, and choosing the distance $a$ smaller than the coherence length $r_0$ will avoid phase unwrapping.

When two beams are overlayed, a tilt can only be applied in one coordinate direction, and a phase difference --- gradient --- can be measured in this
direction. To retrieve gradients in two coordinates several methods are possible: such as overlapping more than two beams, or overlapping perpendicular
polarization measurements and disentangling the polarization planes at the detector. These possibilities will be outlined in Section~\ref{sec:twocoord}. First
we describe the basic principle of a two beam overlap and detection.

\subsection{Formation of the interferogram}

Due to the coherence of the laser wavefronts, the intensity at each position (x,y) follows the equation of a two-beam interference:
\begin{equation}\label{basic_PSI}
I(x,y,\alpha)=I{'}(x,y)+I{''}(x,y)\cos{(\Phi_a(x,y)+\Phi_t(x,y)+\alpha)}
\end{equation}
were $I{'}(x,y)$ is the sum of the two beam intensities or the average intensity and $I{''}(x,y)$ represents the fringe modulation. $\Phi_a(x,y)$ is the phase
difference between the optical path lengths that is accumulated in the atmosphere, and $\Phi_t(x,y)$ denotes the phase difference due to the relative tilt of
the two waves and all components present before traveling through the medium to be probed. $\alpha$ is the global additional phase difference which is
artificially added to the launched beam. Shifting $\alpha$ in steps from pulse to pulse allows $\Phi_a(x,y)$ to be extracted, as will be outlined in
Section~\ref{sec:algorithm}. Referring to Fig.~\ref{fig:basic_principle}, it is obvious that the total number of fringes $n$ resulting from the relative tilt,
over a telescope aperture of size $D$, depends on the angle between the beams expressed here with the scattering height $H$ and the chosen distance $a$:
\begin{equation}
n=\frac{a D}{H \lambda}
\end{equation}
where $\lambda$ is the wavelength of the laser. The phase difference $\Phi_a(x,y)$ is the measurement signal that has to be extracted, and represents the phase
gradients that have to be corrected with the adaptive optics system. The measured amplitude of $\Phi_a$ will depend on the choice of the initial shear of the
laser beams and the relative tilt which is applied.

\begin{figure}
\centering
\includegraphics[width=7cm]{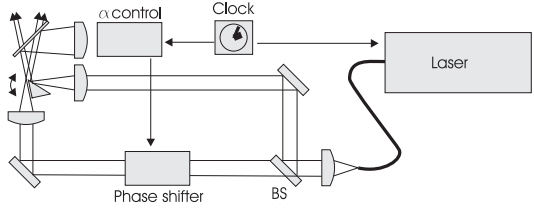}
\caption{Possible arrangement for the beam combiner and phase
  shifter. The incoming laser pulse is split into two beams. In one
  path a --- preferably electro-optic --- phase shifter is
  installed. On combining the two beams, the separation and
tilt can be adjusted. A master clock would control the phase shifting according to the pulse
  sequence which is coming in.
For high accuracy it might be necessary to have a fast control loop stabilizing the phase shift at
  the commanded value. } \label{fig:phaseshifter}
\end{figure}

When the overlapped laser wavefronts traveling upwards are contained in a short duration pulse, a camera which is gated to the time-of-flight for a certain
height can image the pattern created there due to scattering by molecules. Using several consecutive pulses in a very short time and shifting the relative
phase of the wavefronts by a certain angle $\alpha$, the unknown phase difference $\Phi_a$ can be computed with the usual PSI algorithms. When all the
intensities in Eq.~\ref{basic_PSI} are treated as unknown, there are three unknown terms, and thus a minimum of three laser pulses are necessary to retrieve
$\Phi_a$. These pulses have to probe the atmosphere within a timescale in which the phase $\Phi_a$ can be regarded as constant. In the visible, the coherence
time at typical astronomical sites is of order 5\,ms, thus several pulses would need to be launched and the backscatter detected within this time. For a
scattering height of 10\,km the traveling time up and down is only 67\,$\mu s$. But the next pulse can only be launched when there is no contamination from the
first pulse on the second detection, which is approximately the travel time to a height of 30\,km, where the last scattering occurs in the atmosphere, and
back. This is a time of 0.2\,ms, short enough that, including some contingency, three to four pulses could be launched within a millisecond. For a realistic
system, a design optimized for the I-band and longer wavelengths would relax the temporal requirements, and a kHz laser and detection system should be fully
sufficient.

\subsection{Imaging of the pattern}

The backwards imaging of the pattern with a camera will put some constrains on the maximum number of fringes that should be created. Sampling the intensity
distribution, which is described in Eq.~\ref{basic_PSI}, with image points of a certain size will modify the modulation of the detected interferogram. If the
phase change over the pixel is linear, or nearly so, it can be shown easily that integration of Eq.~\ref{basic_PSI} over a range $-\Delta/2, \Delta/2$  leads
to the following expression:
\begin{eqnarray}\label{basic_PSI_samp}
I(x,y,\alpha) & = & I{'}(x,y)  \nonumber \\ & +&{} I{''}(x,y)sinc(\frac{\Delta}{2 \pi }) \cos{(\Phi_a(x,y)+\Phi_t(x,y)+\alpha)}
\end{eqnarray}
with $sinc(\beta)=\frac{sin(\pi\beta)}{\pi\beta}$ and where $\Delta$ is the phase difference over one detection pixel. With the $sinc$ function approaching
zero, the modulation in the interferogram approaches zero as well and the signal to noise is reduced. It will be shown later that up to $\Delta \sim \pi/2$ the
signal to noise is only marginally influenced. The main fringes are present due to the relative tilt of the wavefronts. This means that with a detection camera
having N pixels over the telescope diameter the number of fringes should not be larger than $N/4$, and therefore:
\begin{equation}\label{basic_PSI_N_1}
N=4\frac{a D}{H \lambda}
\end{equation}

To evaluate the approximate number of pixels needed we can choose the distance $a$ to match one pixel, which in general is not a requirement. The measured
$\Phi_a$ then becomes directly the physical phase difference from one pixel to the next and the number of pixels depend on the square root of $1/H\lambda$:

\begin{equation}\label{basic_PSI_N}
N=\frac{2D}{\sqrt{H\lambda}}
\end{equation}
For an 8-m telescope, 532\,nm laser wavelength and 10\,km gating this would require $N=219$ elements; and for 30\,km we find $N=126$. In choosing the actual
number of pixels, the reduction in modulation due to the $sinc$ function has to be traded against the signal to noise due to the photon number incident on one
detection element.

On the way back from the scattering point, the light will be influenced by the phase distortions above the telescope, in a cone that has its base at the pupil
and its tip at position (x,y). To retrieve the information written in the sky, the image of each point must not be distorted too strongly. A simple estimation
of the angular distance of one element in the sky, as seen from the camera on the ground, shows that in general this will not be a serious problem. The angle
$\Theta$ is computed from $H$ and the number of pixels across the telescope:
\begin{equation}\label{basic_PSI_T}
\tan \Theta=\frac{D}{N H}
\end{equation}
and for a pixel number as assumed in Eq.~\ref{basic_PSI_N}, the angular size of one image element is:
\begin{equation}\label{basic_PSI_TN}
\Theta=\arctan\left( \frac{1}{2}\sqrt{\frac{\lambda}{ H}} \right)
\end{equation}
For a 10\,km detection (at 532\,nm) this amounts to $\approx$ 0.8\arcsec, and for a 30\,km detection the angle seen is 0.4\arcsec. These are within the size
range of a seeing limited PSF. Additionally, global atmospheric tip-tilt is completely excluded from the detection due to Fermat's principle. Therefore the
image will be smeared out less than the seeing limit. In closed loop operation with high order AO, the situation will improve since the image will be
sharpened. Larger scattering heights (e.g. sodium layer excitation) will result in a smaller maximum number of pixels; but due to the lack of cone effect there
is no real advantage in going to greater heights.

If the pixel number is chosen in the range as stated here, there is a perfect coincidence with the turbulent structure of the atmosphere: at short laser
wavelengths more fringes are created, thus more detection pixels are required, and the turbulence is sampled on smaller scales. The $\approx$128-pixel limit
for a 30\,km, 532\,nm detection would divide the telescope into subapertures of 6.25\,cm squared, which is just the number required to sample the atmospheric
turbulence on scales of Fried's coherence length in the visible. Of course most projects will not need the high sampling that is required for short wavelength
correction and it should be noted that the method outlined here can be used with fewer detection elements and a lower number of fringes across the telescope.
By changing the number of fringes, the sensitivity can be adapted to the wavelength of correction and could even be chosen on-line to match the ambient
atmospheric conditions.

There is therefore a wide range of possibilities to arrange the gating height to fit the needs of an individual project: multiple imaging over the propagation,
using Rayleigh scattering, or sodium resonance excitation.

\subsection{Phase retrieval algorithms}\label{sec:algorithm}

There are several standard algorithms to calculate $\Phi_a$ from a set of phase shifted interferograms, which can be found in textbooks (e.g. Malacara
\cite{Malacara92}). With the four-step algorithm four interferograms $I_i(x,y)$ are recorded with phaseshifts of
 $\alpha_i=\{0,\frac{\pi}{2},\pi,\frac{3 \pi}{2}\}$ respectively.
The solution to the four equations contains only the measured intensities:
\begin{equation}\label{eq_4step}
\tan(\Phi(x,y))=\frac{I_4-I_2}{I_1-I_3}
\end{equation}
Comparable solutions can be found for three or more steps, e.g. the Hariharan or Carr\'{e} algorithm. Common to all of them is that the resultant phase is
expressed as the arctangent of the relation of measured intensities. Using modulo 2$\pi$ correction, which is implemented in most computing languages as a
two-argument tangent function, unambiguous results are obtained within a phase range of $[-\pi,\pi]$. Larger wavefront phases are usually calculated with the
assumption of a steady function and a process called phase unwrapping.

Since the method proposed in this paper only measures the phase difference between the sheared beams, phase unwrapping due to the sky component $\Phi_a$ will
not be necessary if the shear distance $a$ is chosen to be of order $r_0$. Assuming, as a back of the envelope calculation, a Gaussian distribution of
gradients, the probability for phase differences larger than $\pi$ to occur is then only 0.0017 and decreases rapidly to $5\cdot10^{-5}$ at a distance $a$
chosen to be $0.8 r_0$.

Recalling Eq.~\ref{basic_PSI} the interferogram contains, besides the atmospheric component, the phase term from the tilt $\Phi_t$ of the laser wavefronts. If
the standard PSI algorithms mentioned above are used, the $\Phi_t$ component will always be present in the measurement, steadily increasing over the measured
frame. This causes pixels on the detector to be close to the ambiguity of the tangent function, and with noise present in the measurement, $2\pi$ errors could
easily occur. Removing those errors via phase unwrapping is computationally demanding. To avoid pixels on the wavefront sensor giving a measurement near the
phase ambiguity and therefore being sensitive to noise, we propose to use the phase retrieval algorithm outlined in the following paragraph. The algorithm uses
calibration frames to remove the $\Phi_t$ component from the measurement. With two reference frames and two on-sky measurements a solution is found from
Eq.~\ref{basic_PSI_cal3}. Combining more phase shifts and reference interferograms leads to an optimum phase measurement, explicitly shown for three steps in
Eqs.~\ref{basic_PSI_cal12}-\ref{basic_PSI_calall}. This treatment shifts the measurement for each pixel into the center of the unambiguous phase range and
hence removes any need for phase unwrapping.

If we assume that in advance of the on-sky measurement a set of interferograms is recorded that do not contain the $\Phi_a$ component, we get in total $N_s$
equations with $i=\{ 1,2,3,...N_s\}$ for the $N_s$ different phaseshifts that are applied:
\begin{equation}\label{basic_PSI_cal1}
I_i(x,y,\alpha)=I{'}(x,y)+I{''}(x,y)\cos{(\Phi_a(x,y)+\Phi_t(x,y)+\alpha_i)}
\end{equation}
\begin{equation}\label{basic_PSI_cal2}
I_{ci}(x,y,\alpha)=I{'}(x,y)+I{''}(x,y)\cos{(\Phi_t(x,y)+\alpha_i)}
\end{equation}
$I_{ci}(x,y,\alpha_i)$ are reference frames which are recorded with the same wavefront tilt, and the same phase shifts $\alpha_i$ for which $I_i(x,y,\alpha_i)$
will be recorded on sky. The reference could either be taken by averaging lots of images from the sky, or a camera or special scattering screen could be
inserted in the AO system conjugated to the height at which on-sky images will later be taken.

The combination of Eqs.~\ref{basic_PSI_cal1} and~\ref{basic_PSI_cal2} leads to an interesting solution for every consecutive $i$ and $i+1$. By using
trigonometrical identities the following equation can be found for each similar combination of $I_{i,i+1}$:

\begin{equation}\label{basic_PSI_cal3}
\tan(\frac{\Phi_a}{2})=\frac{I_1-I_{c1}+I_2-I_{c2}}{I_1+I_{c1}-I_2-I_{c2}}\tan(\frac{\alpha_1-\alpha_2}{2})
\end{equation}

$N_s$ laser shots will give $N_s$ solutions for $\Phi_a$ that are independent of the $\Phi_t$ component and contain the difference of the two phase shifts that
have been applied. It is noteworthy that $\Phi_a$ is measured at half angle, which extends the phase range that can be directly detected to $[-\pi,\pi]$,
cancelling the need for a modulo 2$\pi$ correction. In addition, there is no need to fix the phase shifts at a particular value, since they could be measured
separately for the individual shots and just be included into the calculation. Explicitly for three steps the three solutions are:
\begin{equation}\label{basic_PSI_cal12}
\tan(\frac{\Phi_a}{2})_1=\frac{I_1-I_{c1}+I_2-I_{c2}}{I_1+I_{c1}-I_2-I_{c2}}\tan(\frac{\alpha_1-\alpha_2}{2}):=\frac{I_{n1}}{I_{d1}}
\end{equation}
\begin{equation}\label{basic_PSI_cal23}
\tan(\frac{\Phi_a}{2})_2=\frac{I_2-I_{c2}+I_3-I_{c3}}{I_2+I_{c2}-I_3-I_{c3}}\tan(\frac{\alpha_2-\alpha_3}{2}):=\frac{I_{n2}}{I_{d2}}
\end{equation}
\begin{equation}\label{basic_PSI_cal31}
\tan(\frac{\Phi_a}{2})_3=\frac{I_3-I_{c3}+I_1-I_{c1}}{I_3+I_{c3}-I_1-I_{c1}}\tan(\frac{\alpha_3-\alpha_1}{2}):=\frac{I_{n3}}{I_{d3}}
\end{equation}

These equations show that, in principle, two laser shots and proper calibration frames could be enough to evaluate the phase difference. Due to the error
behavior of the formulae above this is not the best choice. Each time the denominator becomes small, large errors occur in the calculation. If a weighted sum
of the three ($N_s$) equations is used, a solution with minimal error behavior can be found, with the power $k$ being an even number:
\begin{equation}\label{basic_PSI_calall}
\tan(\frac{\Phi_a}{2})=\frac{1}{I_{d1}^k+I_{d2}^k+I_{d3}^k}\{ I_{n1}I_{d1}^{k-1}+I_{n2}I_{d2}^{k-1}+I_{n3}I_{d3}^{k-1} \}
\end{equation}
The algorithms will perform best if the $\alpha$ steps equal $2 \pi /N_s$ evenly distributed around a circle. The same principle can be applied for any number
of $\alpha$ steps, and the detection becomes more stable against phase shifting errors. Due to the limited time in which the laser beams have to be launched,
three or four seems to be a good choice.

\subsection{Detection errors}

\begin{figure}
\centering
\includegraphics[width=7.5cm]{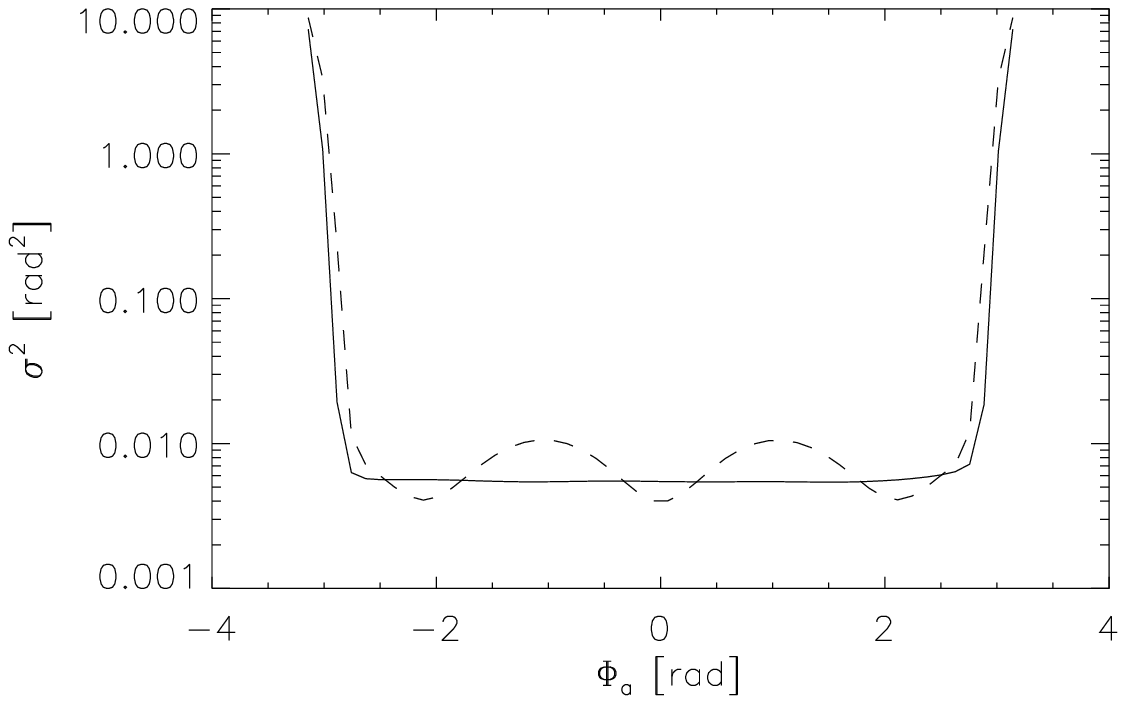}
\includegraphics[width=7.5cm]{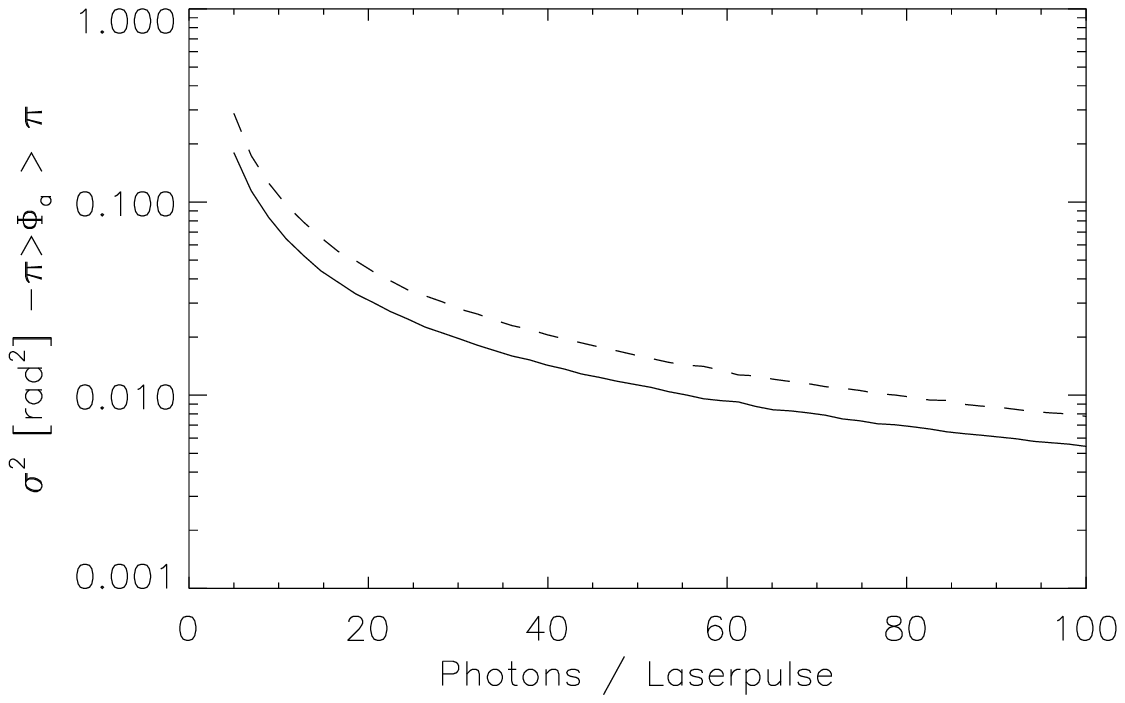}
\includegraphics[width=7.5cm]{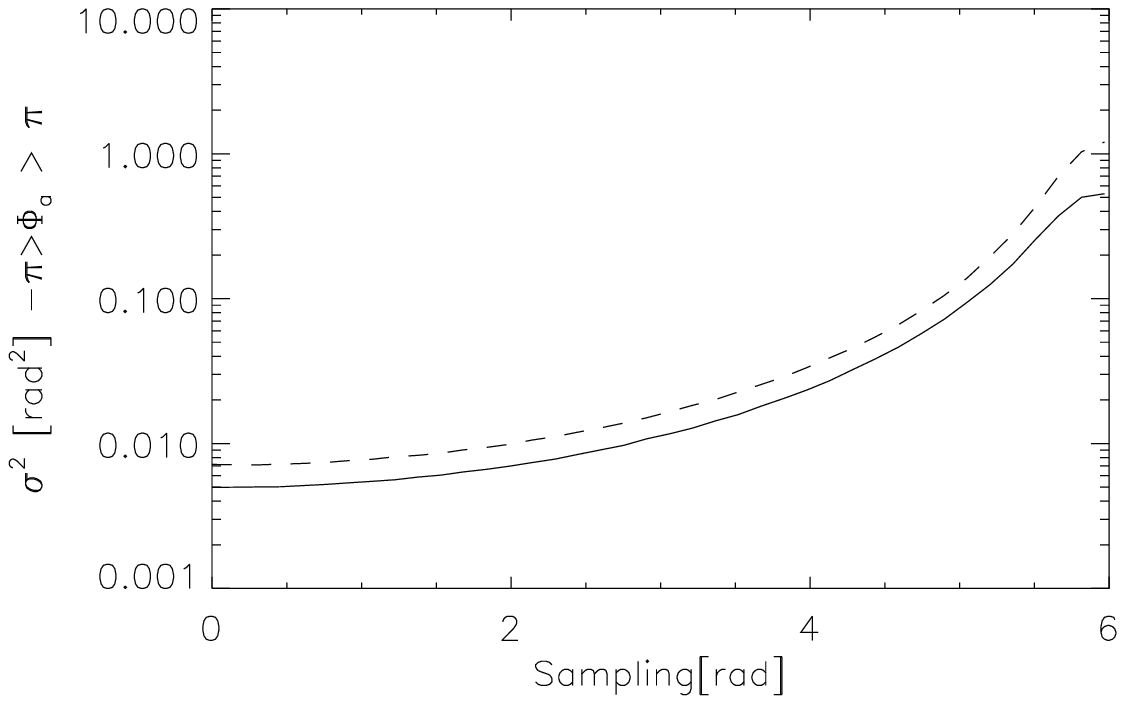}
\caption{Error behavior of PSI algorithms. The top panel shows how the
  best achievable $\sigma^2$ of the gradient measurement behaves over
  the phase range that can be detected.
For this plot 100 photons per laser shot, a detector noise of two
  electrons and a sampling of $\pi/4$ has been assumed.
Near the discontinuities of the tangent $2\pi$, errors occur that
  appear as noise in this plot.
While the noise in the four step algorithm (solid lines) does not depend on the actual phase, the three steps (dashed lines) shows a
  dependence on phase position.
In the middle panel the dependence on photon number of the average variance is shown, including again a detector noise of 2
  electrons and a similar sampling as before.
In the bottom panel, the increase of $\sigma^2$ due to the sampling $\Delta$ from Eq.~\ref{basic_PSI_samp} is drawn, again
  assuming 100 photons, but no detector noise.
Up to a pixel size that spans $\approx \pi/2$, the noise in the
  measurement is only marginally increased.} \label{fig:sig_PSI_alg}
\end{figure}

The number of photons that is necessary per pixel to achieve a good measurement of $\Phi_a$ is a crucial value, since it determines the total laser power
required. Each of the above mentioned algorithms shows a distinct error behavior that depends on the choice of the phase shift $\alpha_i$, and for some
algorithms on the actual phase value under test. In general it has been shown (Bruning \cite{Bruning92}) that intensity fluctuations cause the standard
deviation in the measured phase to be:
\begin{equation}
\sigma=\frac{1}{\sqrt{N}S}
\end{equation}
where $N_s$ is the number of phase steps and $S$ the intensity signal-to-noise ratio. In the case of pure photon noise the standard deviation goes as (Wyant
\cite{Wyant75}):
\begin{equation}
\sigma=\frac{1}{\sqrt{P}}
\end{equation}
where $P$ is the total number of photons detected. Fig.~\ref{fig:sig_PSI_alg} shows the variance of the phase gradients calculated numerically with the
calibrated three- and four-step algorithms. In this figure the photon noise, and a contribution from readout noise and sampling of the interferogram, are
included. Depending on the actual scheme of detection and calibration, either algorithm could be preferable. With a total of 100 photons arriving within the
three or four laser shots on one detector element, the $\sigma^2$ of the on-axis phase gradient can be as low as $\approx0.02rad^2$.

Other error sources that might be special for this method are shot-to-shot variation of laser power and intensity distribution. If this turns out to be a
serious issue the actual intensity profile of the launched beam could be measured with a camera online and taken into account. Incorrect phase shifts are
another possible error source that should be taken into account for a real system. While the limited coherence time of the atmosphere calls for an algorithm
with a low number of phase steps, the stability against phase shifting errors increases with the number of steps taken. Thus for a low number of steps an
accurate phase shifting control should be present. With a fast electro optical phase shifter built into the system, this would be no major difficulty.

\subsection{Gradients in two coordinate directions}\label{sec:twocoord}

\begin{figure}
\centering \includegraphics[width=7cm]{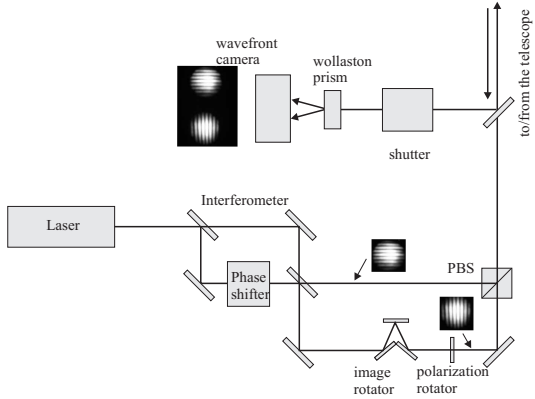} \caption{\small Possible scheme for polarization combining in the
  two-direction gradient measurement. The linearly polarized laser
  beam is split in the interferometer into four overlapping
  beams. While two of these propagate directly to the PBS, the tilt
  and the polarization plane of the other two beams are rotated by
  90$^\circ$ in the image and polarization rotator. Combining all four
  beams is then done in the polarization beam splitter. On the way
  back the planes can be disentangled in front of the wavefront camera
  with a Wollaston prism.}
\label{fig:Polcombine}
\end{figure}

The sheared two beam detection, as mentioned above, only measures the gradient of the wavefront distortions along the direction of the applied tilt. To
retrieve two-dimensional information additional means have to be employed. Apart from the possibility of disentangling x- and y-gradients via temporally
independent measurements, two simultaneous methods will be described here: the use of orthogonal polarization states, and three beam interference. The first
method is straightforward as shown in Fig.~\ref{fig:Polcombine}: one pair of tilted beams is launched linearly polarized. The second pair of beams has its tilt
and the polarization state arranged orthogonally to the first pair. In a polarization beam splitter all the beams are combined and sent to the sky. Since the
back scattering process is fully polarization preserving, the image of the scattered light on the wavefront sensor can again be separated with a polarizing
beam splitter into the desired x- and y-components. The calculation of the wavefront gradients can then follow fully the process described above. While the
laser launching system in this scheme is quite simple and can be made fully polarization preserving, the path through the subsequent adaptive optics and image
de-rotators of the main telescope may cause some trouble due to unwanted mixing of the two polarization states. A careful design and eventually an active
polarization control would be required.

Three or multiple beam interference can also be used to derive information along two directions. If three beams are coherently overlaid, two tilt directions
can be applied. If the central beam is considered as number one, the second beam would be sheared and tilted along an x-direction and the third one along y.
The resultant interference pattern is described by an equation containing six unknowns, and therefore a total of six phase shifts will be required to solve for
$\Phi_{ax}$ and $\Phi_{ay}$. Depending on the speed of the pulsed laser and the detection camera this may be an option, but will take longer than the method
described above. An alternative is the use of Fourier transform methods to derive $\Phi_{ax}$ and $\Phi_{ay}$. Methods exist to calculate gradients from
multiple beam shearing interferograms with Fourier transform techniques (e.g. Primot \cite{Primot93}). Shearing the three beams in the x- and y-direction and
using phase shifts gives easy access to Fourier methods. In the three beam interference case, calculating the differences of two measurements --- one with
$\alpha_x=\alpha_y=0$ and one with $\alpha_x=\alpha_y=\pi$ --- will result in a sum of decoupled cosine functions in x and y. This can be filtered in Fourier
space to disentangle the coordinate directions. A detailed analysis of this method is left for a forthcoming publication.

\subsection{Sensitivity to turbulence at different heights}

The tilt of the coherent waves causes neighboring locations in one layer of the turbulence to be probed. If we consider first one probe of the phase, some
basic relations of the sensing geometry can be found. At ground level, one phase will `see' the atmospheric distortion at the point $(x,y,z_0)$, while the
other traverses the point $(x+a,y,z_0)$. While traveling upwards the distance between the two paths decreases linearly, approaching zero at the height of
detection. The original fringe pattern, formed due to the tilt, is therefore modified by the integral of the optical path length from zero to the detection
height H. This can be regarded as a shear of the phase distortions by the distance $a$ at ground level, and a shear $s$ at height $z$.
\begin{equation}
s=a(H-z)/H
\end{equation}
The measured phase difference $\Phi_a$ is therefore most sensitive to the ground level part of the turbulence, since at higher altitudes nearly the same region
of turbulence is probed. With differential atmospheric phase distortions $\phi_{atm}$ super-imposed onto an on-axis light path, the measured signal is
connected with:
\begin{equation}
\Phi_a=\int_0^H\phi_{atm}(x,y,z)-\phi_{atm}(x+s,y,z)dz
\end{equation}
For a subaperture, with minor high-order distortions, $\Phi_a$ is directly proportional to the average local wavefront gradient $\phi_{atm}'$ multiplied by the
shear $s$:
\begin{equation}
\Phi_a=\int_0^H\phi_{atm}'\cdot s dz
\end{equation}
Due to the insensitivity to turbulent layers near the detection height, low altitude scattering will retrieve mainly the near-ground component. For example, if
scattering at 20\,km height is used, only half of the turbulence strength at 10\,km can be seen. For sodium excitation at 90\,km nearly all the structure of
the turbulence is contained in the measurement. But even so, the remaining 10\% of the distortions from a 10\,km height will prevent diffraction limited
correction from being achieved in the visible.

It should be noted that in comparison with a `standard' laser guide star, the phase disturbance of turbulence at higher altitude is contained within the
measurement, albeit at smaller amplitude. The usual cone under a laser guide star totally misses the structure outside the cone, and does a wrong estimation of
the higher components, due to the projection geometry onto the pupil.

To compare the possible performance of this scheme with a `standard' laser guide star, we adopt as a first estimate the statistical treatment of the phase
error, caused by focal anisoplanatism, as derived by Parenti and Sasiela \cite{Parenti94}:

\begin{equation}
\sigma_{cone}^2=0.5D^{\frac{5}{3}}(\frac{2 \pi}{\lambda})^2 \sec{\xi}[\frac{\mu_{5/3}|_0^H}{H^{5/3}}-0.904\frac{\mu_2|_0^H}{H^2}]
\end{equation}
for the part of the turbulence which is not measured outside the cone, and:

\begin{equation}\label{eq:sigunsample}
\sigma_{above}^2=0.057D^{\frac{5}{3}}(\frac{2 \pi}{\lambda})^2 \sec{\xi}\;\mu_0|_H^\infty
\end{equation}
for the part that is not sensed above the guide star. Here $\mu_n$ is the usual $n$-order moment of the turbulence structure function:
\begin{equation} \label{At:moment}
{\mu_{n}=\int C_{n}^{2}(h)h^{n}dh}
\end{equation}
For PSI sensing with tilted wavefronts, the weakly sensed --- and therefore weakly corrected --- part of the turbulence is proportional to the shear $a-s(z)$.
This assumption will hold as long as the distance $a$ is not much larger than the atmospheric turbulence cells. The variance for the LPSI method due to the
weakly sensed parts is therefore proportional to the first order of $\mu_n$:
\begin{equation}
\sigma_{lpsi}^2=0.057D^{\frac{5}{3}}(\frac{2 \pi}{\lambda})^2 \sec{\xi}\;\frac{1}{H}\;\mu_1|_0^H
\end{equation}
The remaining phase error is always the combination of the unsensed part of the turbulence above the guide star, and a contribution of the insufficient
measurement up to $H$: for single guide stars the part outside the cone, and for LPSI the underestimation of the upper layers. For this estimation a simple
Hufnagel-Valley model of the $C_n^2$ distribution is assumed, as well as a power spectrum as present in Kolmogorov turbulence. The used $C_n^2$ model is:
\begin{equation}\label{eq:cn2}
C_n^2=2.2\cdot 10^{-23} h^{10}e^{-h}+1.10\cdot 10 ^{-16}e^{-\frac{h}{1.5}}+1.7\cdot 10^{-14}e^{-\frac{h}{0.1}}
\end{equation}

\begin{figure}
\centering \includegraphics[width=7cm]{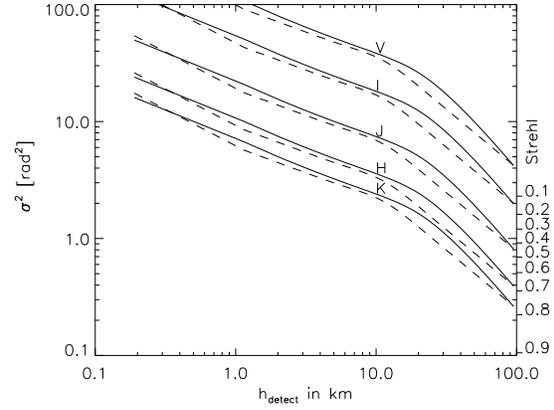} \caption{Variance of the phase against guide star height. Values are calculated for a 8.2-m telescope, with
the $C_n^2$ model from Eq.~\ref{eq:cn2} assumed. Solid lines: single `standard' guide star. Dashed lines: variance of the phase that is corrected with a single
layer LPSI detection. The linear insensitivity to turbulence near the detection height shows an advantage over the quadratic dependence of the single guide
star, especially if the scattering is at medium altitude.} \label{fig:sig_1layPsi}
\end{figure}

In Fig.~\ref{fig:sig_1layPsi} the variance of both possibilities is shown as a function of guide star (i.e. gating) height for an 8-m telescope. The linear
dependence of the weakly sensed areas causes the LPSI method to perform better, especially in the mid-range 10--30\,km detection heights. In total Strehl ratio
(computed with the Mar\'{e}chal approximation), the difference for a telescope at this size is not really significant. For a ground layer adaptive optics
system the linear decrease of sensitivity with height in LPSI could be a real benefit. Compared with single guide star systems, turbulence is never sensed
wrongly or omitted outside a cone. In the following section we will show arrangements that allow for a full turbulence recovery.

\section{Multiple height measurements}

\subsection{Gating Rayleigh scattered light at two heights}

\begin{figure}
\centering \includegraphics[width=4cm]{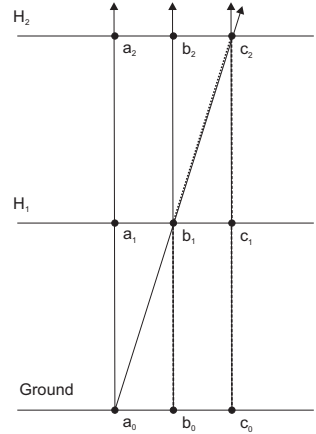} \caption{Arrangement of the tilted waves when multiple scattering from
  the atmosphere is detected.
The camera is gated such that images from at least two different
  heights are retrieved.
By choosing the right pixels, and subtracting them, all turbulence from below the first height can be recovered. The sampled turbulence volume is indicated
with dashed lines.} \label{fig:scheme_2layPsi}
\end{figure}

By extending the proposed scheme to multiple gating heights, more information can be gained on the turbulence volume above the telescope. If we assume a scheme
as shown in Fig.~\ref{fig:scheme_2layPsi}, the wavefronts are tilted such that a `crossing' takes place at the first detection height.
The resultant phase differences are detected at two heights at the points $a_{1,2}, b_{1,2},...$ corresponding to the pixel location of the detector. Taking
now the two measurements of the phase difference at these heights and subtracting the shifted frames from each other, will result in a full on-axis measurement
up to height $H_1$ as can be shown easily: the measurement $\Delta \Phi_{b_1}$ at point $b_1$ is the difference of the collected phase changes along the path
$a_0 \rightarrow b_1$ and $b_0 \rightarrow b_1$:
\begin{equation}
\Delta \Phi_{b_1}=\Phi_{a_0,b_1}-\Phi_{b_0,b_1}
\end{equation}
and the measured phase difference at the point $c_2$ is:
\begin{eqnarray}
\Delta \Phi_{c_2}& =& \Phi_{a_0,c_2}-\Phi_{c_0,c_2}\nonumber \\
&=& \Phi_{a_0,b_1}+\Phi_{b_1,c_2}-\Phi_{c_0,c_1}-\Phi_{c_1,c_2}
\end{eqnarray}
Calculating the difference of the shifted frames results in:
\begin{equation}
\Delta \Phi_{c_2}-\Delta \Phi_{b_1}=\Phi_{b_0,b_1}-\Phi_{c_0,c_1}+\Phi_{b_1,c_2}-\Phi_{c_1,c_2}
\end{equation}
The term $\Phi_{b_1,c_2}-\Phi_{c_1,c_2}$ describes the path from $H_1$ to $H_2$. This measurement will show the decreasing sensitivity with increasing height
to turbulent structures. The first term $\Phi_{b_0,b_1}-\Phi_{c_0,c_1}$ is the description of the full on-axis gradient in the height range from ground level
to the first detection. The sampled path in the atmosphere is indicated in Fig.~\ref{fig:scheme_2layPsi} with dashed lines. Similar relations can be derived
for any use of two consecutive detections.

\begin{figure}
\centering \includegraphics[width=7cm]{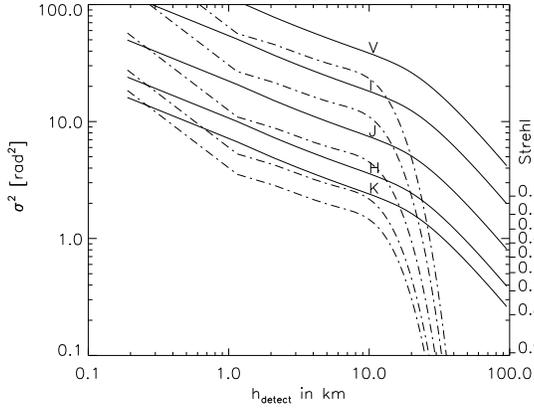} \caption{Solid lines: single `standard' guide star. Dashed lines: Variance of the phase that is corrected
with a double layer LPSI detection in dependence of scattering height h of the upper
  detection.
The lower scattering plane is always chosen to be at half height of the upper one. Values are calculated as before for a 8.2-m telescope, with the
  $C_n^2$ model from Eq.~\ref{eq:cn2} assumed. The linear insensitivity to turbulence near the detection height
  now only shows an effect on the part of turbulence which is above
  h/2. At a detection height
above 25\,km the phase variance in all bands could be reduced below
  0.2 $rad^2$. } \label{fig:sig_2layPsi}
\end{figure}

If the measured gradients could be fully turned into a correction of the disturbed wavefronts, a system performance as shown in Fig.~\ref{fig:sig_2layPsi}
would result. For this calculation the $C_n^2$ model above has been used again. The detection height of the lower gating was assumed to be half of the upper
height. Comparing again with a `standard' single guide star, the system described here would give a much better performance, even at low scattering altitudes.
If the upper detection is above $\approx$25\,km, the on-axis wavefront is sampled completely, so that even in the visible the remaining on-axis $\sigma^2$ is
below 0.2$rad^2$.

There are other possibilities to arrange gating heights and more detections. With several measurements the turbulent structure could be resolved in small
height slices, leading to the possibility of assigning conjugated DMs to each of them; which despite the extra effort, is worth the gain.

There may be some practical problems with the schematics described above. Since we rely on the fact that the subtraction of the two detections takes place in
neighboring pixels, both should be made with the same number of pixels across the telescope aperture. Since scattering from the upper detection results in a
much lower photon flux and also the turbulent structure at larger height showing a larger coherence length, a sampling with larger subapertures would be
desirable. Scaling low resolution upper layer data onto the highly resolved low altitude scattering might be possible, but would probably lower the performance
of the system.

As shown earlier, the sampling limit at a 30\,km detection is around $128\times128$~pixels for a proper measurement at an 8-m telescope. If the same number is
used for the lower detection, the subaperture size is $\approx$6\,cm. This would be surely sufficient to correct in visible wavelengths and to gain high strehl
in the near infrared. The only limit remaining is then the number of photons that can be produced in the scattering process. This number will be discussed
later on.

\subsection{Multiconjugate correction with two beams}

The possibility shown above allows a full recovery of the turbulent structure of the atmosphere. If deformable mirrors with a large number of actuators are
available, a correction can lead to high strehl ratios even in the visible. Nevertheless, the measurement performed with the above system is completely on-axis
and will thus lead to the usual effects of angular anisoplanatism. In the visible the isoplanatic patch is rather small so only a small field of view can be
corrected. By using multiple deformable mirrors conjugated to different heights, the field can be extended.

\begin{figure}
\centering \includegraphics[width=7cm]{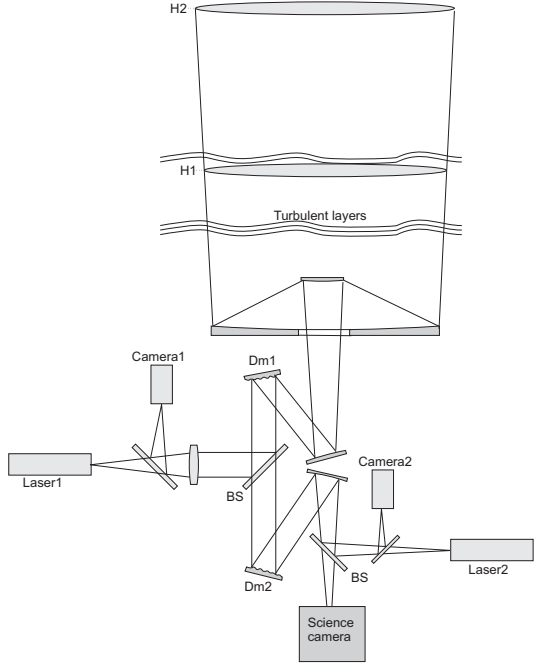} \caption{Schematic arrangement of a multiconjugate correction with
  lasers and LPSI.
Laser 1 is directed over the first deformable mirror, and the
  backscatter detected at lower altitude.
Laser 2, operating at a different color, is directed first over the second deformable mirror, which is conjugated to a high altitude
  turbulence.
After passing a dichroic splitter it joins the beam path of laser 1 and the backscatter is detected from a higher altitude $H_2$.} \label{fig:multiconpsi}
\end{figure}

The LPSI method is almost perfectly suited for this application. In Fig.~\ref{fig:multiconpsi} a possible system is shown. A first laser beam (consisting of
overlaid coherent wavefronts) is directed over the first deformable mirror to the telescope, expanded and launched over the full aperture as described
previously. This beam is detected at lower altitude and steers the first deformable mirror in a closed loop scheme. A second laser beam is then injected before
that, hits the second DM and then joins the beam path of the first laser at a beam splitter. To enable the bams to be properly separated, the two lasers could
operate at different colors. The second laser is detected at higher altitude, and DM number two is controlled, independently from loop number one, by this
signal. This leads to a correction scheme that is computationally not demanding, somehow similar to the `layer oriented' concept proposed by Ragazzoni
\cite{Ragazzoni00}. Treating the process mathematically, the lasers sample differential phases $\phi_{atm}'$ according to the shear $s$:

\begin{equation}
\begin{array}{lcl}
\Phi_{a1}&=&\int_0^{H1}\phi_{atm}'\frac{a(H_1-z)}{H_1}dz\\
\Phi_{a2}&=&\int_0^{H2}\phi_{atm}'\frac{a(H_2-z)}{H_2}dz\\
\end{array}
\end{equation}
Since beam number two is sent via the first DM to the sky, the actual phase which is sensed with it is the difference of $\Phi_{a2}$ and $\Phi_{a1}$, and the
laser wavefront leaves the telescope already corrected for $\Phi_{a1}$. This leads to following expression when setting additionally $H_2=M \cdot H_1$:
\begin{equation}
\Phi_{a2}-\Phi_{a1}=\frac{Ma-a}{M H_1}\int_0^{H1}\phi_{atm}'z dz+\frac{a}{H_2}\int_{H_1}^{H2}\phi_{atm}'(H_2-z) dz
\end{equation}
How strongly this signal is applied to the DM is a function of the conjugated position, where it is located. The total correction over both DMs is therefore
written as:
\begin{equation}
\Phi_c=\Phi_{a1}+Q(\Phi_{a2}-\Phi_{a1})
\end{equation}
with $Q$ being the wavefront shear at the location of DM\,2. If $Q$ is chosen to be:
\begin{equation}
Q=\frac{M}{M-1}
\end{equation}
all $z$-dependent terms in above equation cancel up to $H_1$. The remaining terms are:
\begin{equation}
\Phi_c=a\int_{0}^{H_1}\phi_{atm}'dz+Q\frac{a}{H_2}\int_{H_1}^{H_2}\phi_{atm}'(H_2-z)dz
\end{equation}
The first term describes the complete sampling of all gradients up to height $H_1$. The second term shows the linear decrease in sensitivity from $H_1$ to
$H_2$. The statistical treatment leads to a picture which is similar to the one shown in Fig.~\ref{fig:sig_2layPsi}. Sensing again at 10 and 20\,km the
turbulence is sampled nearly completely --- leading to possible high strehl ratios in all wavelength bands. It is noteworthy that in this scheme there is no
computational subtraction of the two detection heights involved. The difference in the above equations results from the optical setup. Therefore it should be
easily possible to sample the second laser with fewer subapertures. And since the control loops of the two DMs can be kept separate, sampling at different
speeds of both lasers is easily possible. This would greatly relax the requirements for power of the second laser.

It should be noted that one is completely free to choose the type of scattering involved. Both upper and lower altitude detection could be Rayleigh-type, or
the upper scatter process could take place in the earth's sodium layer.

Expanding the scheme outlined above allows one to gain additional information: when the lasers are launched slightly defocused, higher altitude layers are
sampled outside the pupil of the telescope. Applying this information to larger deformable mirrors would increase the off-axis correction. The phase retrieval
will not suffer from such an expanded beam, since only relative phase differences are measured against a calibration. The achievable corrected field depends on
the structure of the turbulence in the atmosphere, and how well the second DM is conjugated to a strong high altitude layer. A detailed analysis of this topic
is out of the scope of this paper and will be given in a forthcoming publication.

\section{Photon numbers and laser requirements}

The number of photons required to achieve a good phase measurement is no different to the usual calculation for laser guide stars. The sensing geometry is just
inverted: with a standard guide star the telescope pupil is divided into sub-apertures; with the LPSI setup, the full aperture of the telescope is used for the
backwards imaging, but the pattern is divided by the number of image elements of the camera. In the following discussion, the photon number is counted to be
the average over the fringes.

\begin{figure}
\centering
\includegraphics[width=6cm]{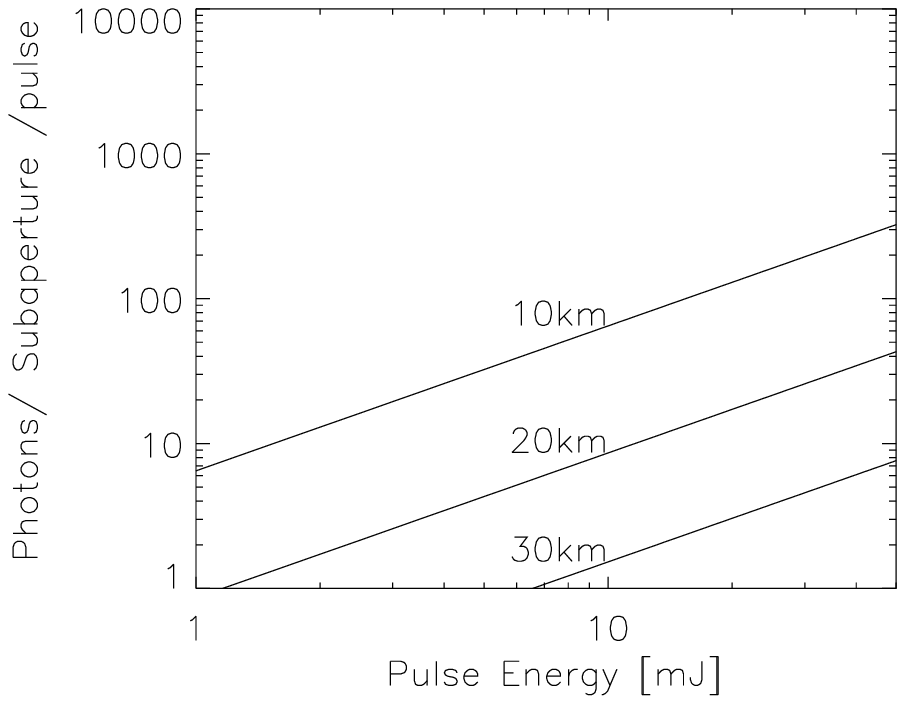}
\includegraphics[width=6cm]{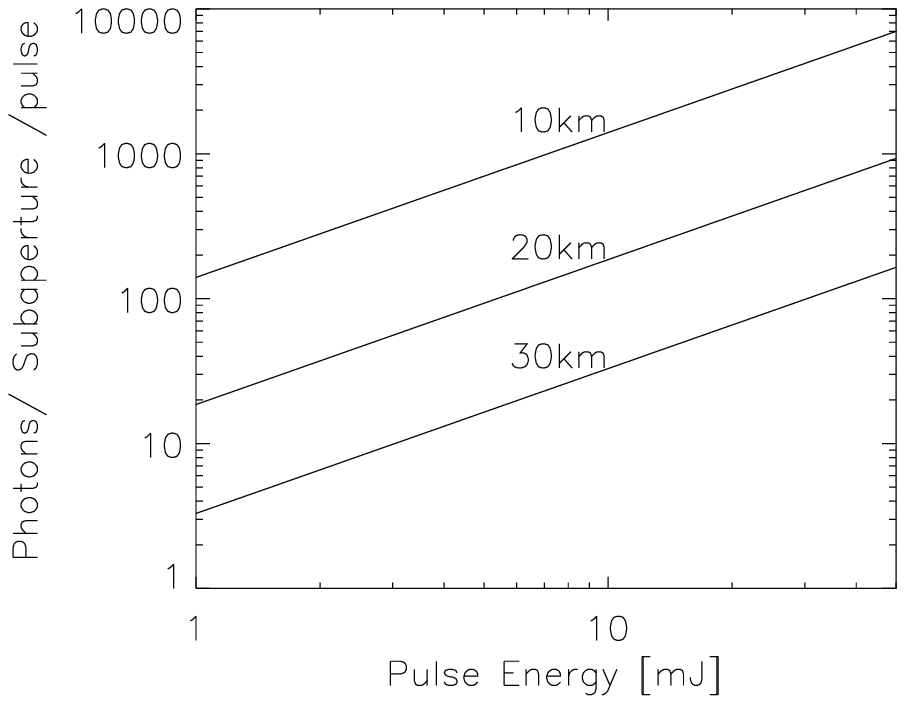}
\caption{Photons collected by a single subaperture during one laser pulse with Rayleigh scattering in the atmosphere. The subaperture size is assumed to be
15\,cm and the overall efficiency is taken to be 25\%. Left panel: the situation for a gating $\Delta H$ of 90\,m at 10\,km and 280\,m at 30\,km are shown.
Right panel: photons collected if a refocussing device were used. Refocussing would take place over 20\% of the gating height. The laser wavelength is assumed
to be 527\,nm.} \label{fig:rayleigh1}
\end{figure}

For Rayleigh scattering the photon number $N_{ph}$ is calculated from the standard theory by (see for instance van de Hulst \cite{vandehulst81}):
\begin{equation}\label{Eq:Rayleigh_photons}
N_{ph}=\eta\frac{E D^2 \rho (H) \frac{d\sigma}{d\Omega}\Delta H }{N^2 \gamma H^2}
\end{equation}
$E$ being the energy of the pulse, $\gamma$ the energy of one photon, $\rho (H)$ the number density of molecules, $H$ the scattering altitude, $D$ the diameter
of the telescope, $N$ the linear number of pixel across the aperture and $\Delta H$ the length over which the scattering is sampled. $\eta$ denotes the overall
efficiency of a round trip from laser to receiver. $\frac{d\sigma(\alpha)}{d\Omega}$ is the scattering coefficient per unit solid angle $\Omega$, in the
direction $\alpha$ with the angle $\beta$ between scattering plane and polarization direction:
\begin{equation}
\frac{d\sigma(\alpha)}{d\Omega}=4 \pi^2\frac{(n_0-1)^2}{N_0^2\lambda^4}(\cos^2(\beta)\cos^2(\alpha)+\sin^2(\beta))
\end{equation}
$n_0$ denotes the refractive index and $N_0$ the number of particles at sea level. Using the usual density models of the atmosphere $N_{ph}$ can be computed.
For the plots in Fig~\ref{fig:rayleigh1} the USSA-1962 model from McCartney \cite{McCartney76} has been used. It should be emphasized that the number of
photons incident on one detector element for a given laser power does not depend on telescope size but only on the chosen size of the subaperure $D/N$.

With `normal' gating of the wavefront camera, the scattering altitude $\Delta H$ has to be kept very small to avoid a smearing of the image while the laser
pulse is traveling this distance. The angular size $\delta$ of $\Delta H$ as seen from the outermost areas of the detector is:
\begin{equation}\label{eq:rangeangle}
\tan \delta=\frac{D \Delta H}{H^2-H\Delta H +D^2}
\end{equation}
which is equivalent to the focal depth of the camera. If we allow this angle to extend over half a pixel and combine Eq.~\ref{eq:rangeangle} with
Eq.~\ref{basic_PSI_T} we find:
\begin{equation}\label{eq:rangegate}
\Delta H=\frac{1}{2}\frac{H^2+ND^2}{H(1+N)}
\end{equation}
To avoid intensity being recorded in the wrong subapertures the possible gating height for an 8-m telescope at 10\,km is only $\approx$90\,m, increasing to
280\,m at 30\,km. This is a fairly small volume out of which scattered light can be collected, and decreases even more for a larger telescope. This situation
is reflected in the upper plot of  Fig.~\ref{fig:rayleigh1}.
For a measurement with 15\,cm subapertures, the photon flux from a 10\,km detection height is just sufficient with a 20\,mJ laser (20\,W average power at
1\,kHz repetition rate), when assuming an overall efficiency of $\eta$=0.25 and a laser wavelength of 527\,nm. A 30\,km detection would require larger
subapertures and very high pulse energy.

The situation changes when using a device that is capable of keeping the image in focus during a part of the travel time. Proposed by Angel \cite{Angel00},
dynamic re-focusing would collect many more photons. Depending on the time during which the image could be stabilized, the power demands for the laser
decreases rapidly. This is shown in the left panel of Fig.~\ref{fig:rayleigh1}. If the range over which photons are collected is increased to 20\% of the
scattering height a 20\,mJ laser at 527\,nm would be sufficient for a 30\,km detection with 15\,cm subapertures. For the case of an extremely large telescope
in the size range of 30\,m or more, a refocussing device would be mandatory, since the focal depth of normal gating will decrease to very small values.

Use of UV-lasers would be another possibility to increase the number of collected photons, due to the $\lambda^{-4}$ dependence of the scattering coefficient.
This has to be balanced against the quantum efficiency of available detectors, the optical efficiency of the system at UV-wavelengths and the available laser
power at the desired repetition rate.

Sodium layer excitation might be an option to collect photon numbers comparable to a 30\,km height detection more easily. Due to the low repetition rate that
is needed ($\approx$1\,kHz, compared to usual 30--60\,kHz for sodium guide star creation) saturation would become a serious issue in a small laser spot.
Fortunately the area which is illuminated is very large. For an 8-m aperture we calculate that saturation will not play a major role for a linewidth of 1\,GHz
with up to 20\,W laser power. For extremely large telescopes the power density will decrease even further.

Considerations concerning the beam quality of the laser are totally relaxed in the proposed system, compared with `standard' guide stars. For the usual sodium
layer guide stars a beam quality of $M^2<1.3$ is usually mandatory. With the splitting and coherent overlap of the laser wavefronts and the calibrated LPSI
algorithms proposed here, the initial wavefront distortions will cancel out completely and spot size is not an issue. In fact a typical top-hat beam profile of
a powerful pulsed laser would be even better than a Gaussian profile from a $TEM_{00}$ beam since the latter would under-illuminate the edges of the pupil.
Lasers that fulfil these requirements are nowadays standard products, available at reasonable cost.

\section{Extension to measure tip-tilt from the laser}

\begin{figure}
\centering
\includegraphics[width=5cm]{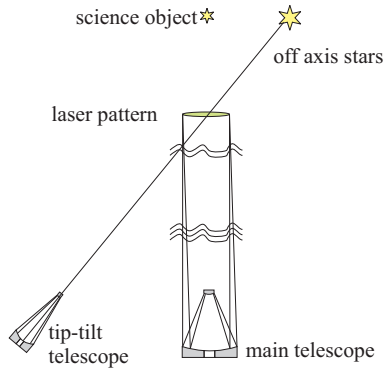}
\caption{Schematic showing how to measure tip-tilt above the main
  telescope by comparing the motion of the laser pattern as seen from
  a side telescope to that of an off-axis bright star.}
\label{fig:tiptilt}
\end{figure}

As already suggested by Ragazzoni \cite{Ragazzoni95}, the possibility exists to expand the area in which a suitable tip-tilt star can be found by using a small
auxiliary telescope. It is worth mentioning that this ideally suited to the laser launching method proposed in this paper. From the laser pattern itself and
the use of the main telescope for detection, the tip-tilt is perfectly invisible due to Fermat's principle. Therefore an additional reference star has to be
used, as in any `standard' LGS system. This puts constraints onto the sky coverage of the system. Some areas which are of particular interest, such as deep
fields, are selected due to the absence of bright stars. The use of the whole telescope as a laser projector, as proposed here, facilitates the use of an
auxiliary telescope for tip-tilt sensing. The principle is illustrated in Fig.~\ref{fig:tiptilt}. With the small telescope capable of moving around over some
area, a far off-axis star can be selected to be along the line of sight of the laser pattern projected onto the sky from the main telescope. At a height above
all main turbulent layers of $\sim$20--30\,km, the global movement of the scattered laser pattern will represent the tilt component of the wavefront, since the
laser beam has probed the full pupil size of the telescope. Subtracting the motion of the star from that of the laser patch, as seen from the side telescope,
will result in a measure of the real tip-tilt along the science beam. For telescopes of smaller size the gain in sky coverage can be great; but for the case of
an extremely large telescope the gain in sky coverage with the external tip-tilt tracking might only be marginal, since the global wavefront slope can be
measured with very faint natural stars.

\section{Conclusions}

We have proposed a method to measure and correct atmospheric distortions for astronomical telescopes using a laser to probe the turbulence. The use of laser
wavefronts over the full aperture and the technique of phase shifting interferometry allows the distortions to be measured and corrected without suffering from
the cone effect. The system outlined is therefore applicable to any size of telescope including extremely large apertures. While the number of actuators on the
deformable mirror and the number of detection pixels on the wavefront sensor have to be adopted accordingly, the laser system itself would not change with
telescope size. We have shown that algorithms for the interferometric phase retrieval exist and allow for a calibration of the measurement. The basic scheme
outlined, involving a single detection of the scattered light, gives easy access to ground layer adaptive optics correction. By using of two deformable mirrors
and two laser beams, a multiconjugate correction can be achieved. The necessary lasers are commercially available, and the power demands can be reduced with a
dynamical refocusing device. The total system can be built relatively compactly, and the laser plus adaptive optics would plug into the telescope as one
instrument. This removes the need for separate laser projectors, simplifying the implementation. The two examples given here of how to arrange the collection
of the interferograms and the correction schemes, can easily be modified to fit the needs of an individual project. While the basic feasibility has been shown
in this paper, further study is required in order to address detailed questions arising from the implementation or extension of such a system. Examples are
field-of-view considerations in the multiconjugate case, studies for stray light suppression onto the science camera, and multi-color sensing for phase range
extension.

\end{document}